# Radial gradient of superionic hydrogen in Earth's inner core

Zepeng Wu[1,#], Liangrui Wei[1,#], Chen Gao[1], Shunqing Wu[1], Renata M. Wentzcovitch[2-4], Yang Sun[1,*]

[1]Department of Physics, Xiamen University, Xiamen 361005, China
[2]Department of Applied Physics and Applied Mathematics, Columbia University, New York, NY 10027, USA
[3]Department of Earth and Environmental Sciences, Columbia University, New York, NY 10027, USA
[4]Lamont–Doherty Earth Observatory, Columbia University, Palisades, NY 10964, USA

Hydrogen is considered a key light element in Earth's core, yet the thermodynamics of its superionic phase and its distribution in the inner core remain unclear. Here, we compute ab initio Gibbs free energies for liquid and superionic hcp and bcc Fe–H phases and construct the superionic–liquid phase diagram over pressure–temperature conditions relevant to the Earth's inner core. We find that phase diagrams at different inner-core pressures collapse when temperatures are scaled by the melting temperature of pure iron, indicating that solid–liquid partitioning is controlled primarily by a reduced temperature relative to iron melting and is weakly sensitive to pressure. This scaling relation further reconciles previously reported discrepancies in partition coefficients among theoretical studies and yields good agreement with available experimental data at low pressures. By applying thermochemical constraints, our free-energy results reveal a radial hydrogen gradient within the inner core. These results demonstrate that compositional gradients of superionic hydrogen in the inner core emerge naturally from equilibrium thermodynamics and suggest a general mechanism governing the depth-dependent distribution of light elements within Earth's inner core.

## INTRODUCTION

The Earth's core consists of a solid inner core (IC) surrounded by a liquid outer core (OC). Both regions are dominated by iron, but additional light elements are required to explain the core's density deficit compared to pure Fe (1–4). The composition and distribution of these light elements are therefore central to understanding the structure and evolution of the core (1, 5, 6). While the OC is generally found to be well mixed, the IC exhibits a more complex internal structure, including pronounced seismic anisotropy and heterogeneities (5–7). These anomalous seismic features have been attributed to the elastic properties of core-forming materials and preferred lattice orientations in the IC (8–10). Based on comparisons of the elastic properties of Fe alloys with different compositions, recent experimental studies have further suggested that radial compositional gradients within the IC may provide improved explanations for the observed seismic anisotropy (11, 12). However, the origin of these gradients remains poorly understood.

Hydrogen is considered a key light element in the core (1, 3, 5), owing to its high cosmic abundance and strong siderophile behavior under high pressure–temperature conditions (13–15). Recent work has suggested that hydrogen enters a superionic state in the hexagonal close-packed (hcp) structure of Fe under IC conditions, exhibiting liquid-like diffusion within the crystalline lattice (16, 17). The superionic state is fundamentally distinct from a conventional solid and can affect the thermodynamic properties of different phases (18). Although the IC is commonly assumed to adopt the hcp structure (19, 20), several elements have been shown to stabilize the body-centered cubic (bcc) phase at core conditions (21–25). Whether hydrogen can also exist as a superionic species in the bcc lattice, and how it influences the competition between hcp and bcc phases in the IC, remains unclear.

Given the potentially large abundance of hydrogen in the core, its partitioning between solid and liquid iron has been extensively investigated in order to constrain hydrogen concentrations in both the IC and OC. High-pressure experiments by Oka et al. and Hikosaka et al. reported solid–liquid partition coefficients of hydrogen, $D_{\mathrm{H}}^{\mathrm{S/L}}$, of approximately 0.7–0.8 at (1900 K, 45 GPa) (26) and (2100 K, 56 GPa) (27), respectively. Because direct experiments at inner-core boundary (ICB) conditions remain challenging, recent theoretical studies have used machine-learning interatomic potentials (MLPs) to estimate $D_{\mathrm{H}}^{\mathrm{S/L}}$ at higher pressures and temperatures. These studies, however, report largely varying results: Yuan and Steinle-Neumann (28) obtained values of 0.50–0.62 under ICB conditions, Liu and Jing (29) reported smaller values of 0.29–0.46 at ICB pressure, and Zhang et al. (30) suggested a nearly constant value of ~0.56 across IC conditions. The origin of these discrepancies is unclear, as these simulations employed different MLPs and different protocols for computing the partition coefficient. Such inconsistencies hinder further development of compositional models for multicomponent Fe–light-element alloys in the core.

Both element partitioning and phase competition are essentially governed by the thermodynamic properties of the phases. A comprehensive understanding of hydrogen in the

[#]These authors contributed equally.
[*]Email: yangsun@xmu.edu.cn

IC therefore requires direct access to the Gibbs free energies of superionic and liquid Fe–H phases, from which phase diagrams and equilibrium compositions can be determined. In this work, we employ a recently developed *ab initio* framework to compute Gibbs free energies and construct phase diagrams for superionic and liquid Fe–H under core conditions (25). This framework has been demonstrated to accurately describe the superionic Fe–O system at extreme pressures and temperatures (25). Using this approach, we investigate the phase competition among superionic hcp, superionic bcc, and liquid Fe–H, assess the pressure dependence of the superionic–liquid phase diagram, and establish thermodynamic constraints on hydrogen distribution from ICB to the center of the IC. We aim to provide interpretations of previously reported discrepancies in $D_{\mathrm{H}}^{\mathrm{S/L}}$ and to elucidate the thermodynamic origin of a possible radial chemical gradient in the IC.

## RESULTS

### A. Superionic-liquid phase diagram

We first simulate Fe-H alloy in the hcp, bcc, as well as liquid phase with *ab initio* molecular dynamics (AIMD). Hydrogen exhibits a superionic state in both hcp and bcc lattices, where its diffusion coefficient is comparable to that in the liquid phase (*SI Appendix*, Fig. S1). Figure 1*A* shows the partial pair correlation functions (PPCF) of $Fe_{85}H_{15}$ in superionic hcp, superionic bcc, and liquid phases at 323 GPa and 5500 K. The Fe–H PPCF shows a nearest-neighbor peak at ~1.5 Å in all phases, indicating strong Fe–H bonding that is insensitive to the Fe phase. The H–H PPCF rapidly approaches unity, indicating weak H-H bonding in all three phases, consistent with the liquid-like, superionic behavior of hydrogen within the crystalline lattices.

To study the superionic–liquid equilibrium at a large scale, we develop an interatomic potential for Fe-H system under core conditions, capable of simulating the superionic state in both hcp and bcc phases (see Method and *SI Appendix*, Note S1). This potential enables superionic–liquid coexistence simulations (Fig. 1*B*), and classical-to-*ab initio* thermodynamic integration calculations (31) to obtain the *ab initio* Gibbs free energies of the superionic and liquid phases at 323 GPa over a wide range of temperatures (see *SI Appendix,* Note S2 and Figs. S3–S9). The Gibbs free energies at 5500 K for superionic hcp, superionic bcc, and the liquid are shown as an example in Fig. 1*C*. For all hydrogen compositions investigated here, the bcc phase consistently exhibits a higher free energy than the hcp phase. Thus, similar to pure Fe (20), the bcc phase remains metastable, while the hcp phase is thermodynamically stable with hydrogen contents up to ~20 at.%. Therefore, only the superionic hcp phase can coexist in equilibrium with the liquid in Fe–H system. The uncertainties in the free-energy calculations are 1–2 meV/atom for the superionic and liquid phases (see *SI Appendix*, Note S4). We determine the solidus and liquidus curves by constructing the common tangent between the liquid and superionic hcp free-energy curves, as illustrated for 5500 K in Fig. 1*C*. Repeating this procedure at different temperatures (*SI Appendix*, Fig. S9) yields the superionic–liquid phase diagram at 323 GPa shown in Fig. 1*D*. In contrast to the small solubility of superionic oxygen (~0.4 at.%) at 5800 K and 323 GPa (25), hydrogen exhibits a much higher solubility, reaching 10 at.% under the same conditions.

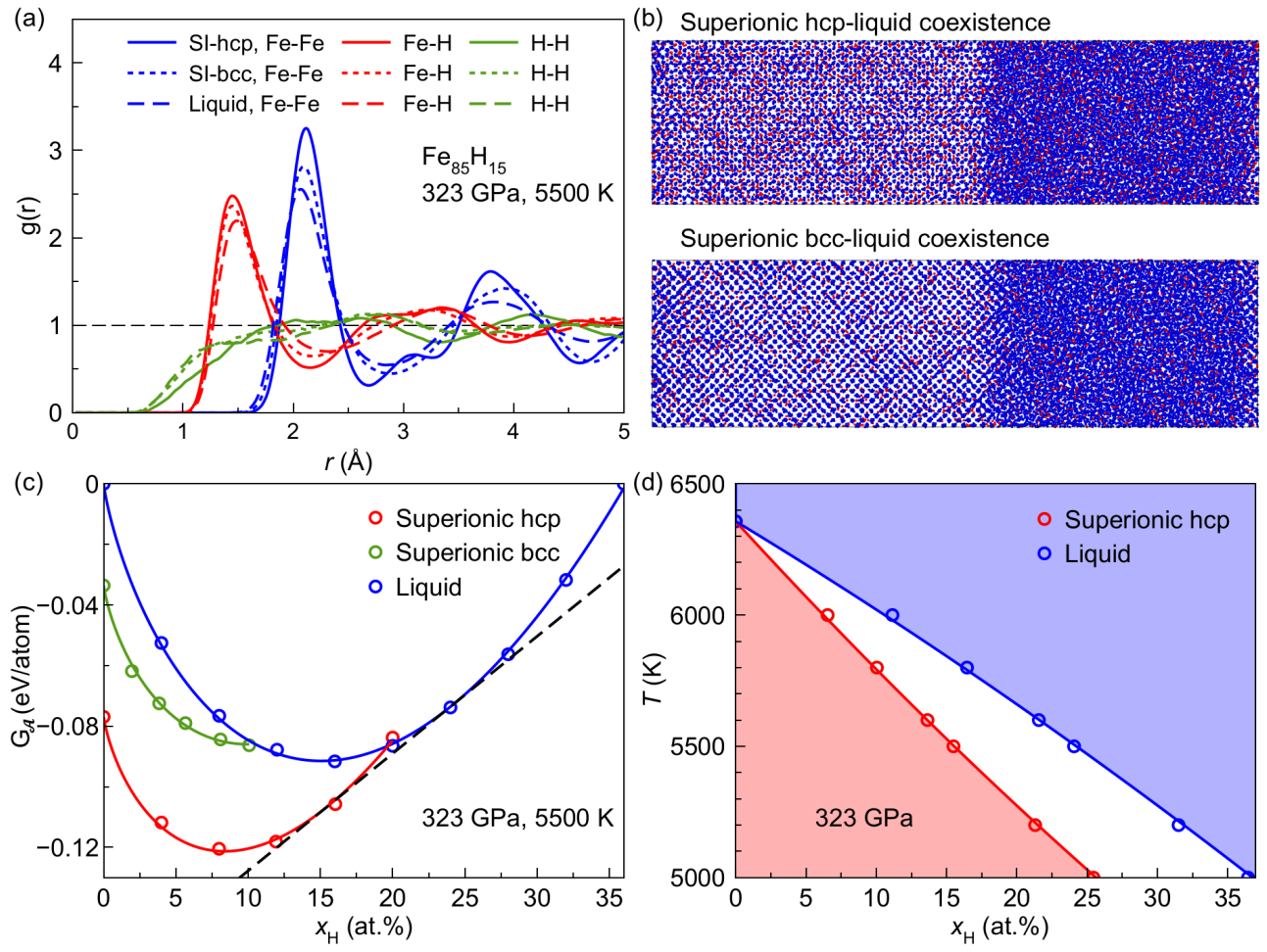

**Figure 1. *Ab initio* superionic-liquid phase diagrams of $Fe_{1-x}H_x$ at 323 GPa. A.** Partial pair correlation functions (PPCF) for superionic hcp, superionic bcc, and liquid $Fe_{85}H_{15}$ at 323 GPa and 5500 K from AIMD. **B.** Superionic hcp–liquid and bcc–liquid coexistence simulations. Blue and red dots represent iron and hydrogen atoms, respectively. **C.** *Ab initio* Gibbs free energies of the liquid, superionic hcp, and superionic bcc $Fe_{1-x}H_x$ phases at 5500 K and 323 GPa, referenced to the liquid free energy at $x_H$=0 at.% and $x_H$=36 at.% for clarity. The black dashed line denotes the common tangent between the liquid and superionic hcp free-energy curves. **D.** *Ab initio* superionic-liquid phase diagrams of $Fe_{1-x}H_x$ at 323 GPa. The red and blue represent the superionic hcp-stable and liquid-stable regions, respectively.

We perform a large number of free-energy calculations for superionic hcp, superionic bcc, and liquid Fe–H phases with different compositions to cover a broad pressure–temperature range relevant to Earth's IC. Figure 2*A* shows an example of the free energy calculation for superionic hcp with $x_H$=20 at.%. Starting from 6000 K and 323 GPa, we compute the Gibbs free energy along the isothermal path using the thermodynamic integration of pressure–volume data at 6000 K over 323–360 GPa obtained from AIMD simulations (see Methods and *SI Appendix,* Note S3 and Figs. S10–S15). Then at 360 GPa, we employ the Gibbs–Helmholtz relation to compute the free energy along the isobaric path. In this manner, Gibbs free energies are computed over the full pressure–temperature space spanning 323–360 GPa and 5000–6800 K for superionic hcp, superionic bcc, and liquid phases of different compositions. The compositional dependence of these free energies is fitted using the Redlich-Kister (RK) expansion (32).

The resulting free-energy data enables analysis of phase competition over a wide range of *P*–*T*–*x* conditions. We first compare the free energies of the superionic hcp and bcc phases. While the superionic hcp phase is stable across most of the investigated *P*–*T*–*x* space, sufficiently high temperatures and hydrogen contents can stabilize the superionic bcc phase. For example, in Fig. 2*B*, the superionic bcc phase becomes more stable than the hcp phase when the temperature exceeds 6400 K and the hydrogen content exceeds 20 at.% at 360 GPa. However, when the liquid phase is taken into account, the stability field of superionic bcc is unlikely to be realized, as the hcp–bcc phase boundary lies above the liquidus line in Fig. 2*B*. Figure 2*C* presents the superionic–liquid phase diagram between 323 and 360 GPa, where only the hcp phase appears as the stable solid. Both the solidus and liquidus lines shift systematically to higher temperatures with increasing pressure. In Fig. 2*D*, we find that when temperature is normalized by the melting temperature of pure Fe at each pressure, using the reduced temperature $\delta T = (T_{m,\mathrm{Fe}} - T)/T_{m,\mathrm{Fe}}$, the solidus and liquidus lines at 323 and 360 GPa collapse onto nearly identical curves. This scaling behavior indicates that pressure primarily rescales the absolute free-energy scale of pure Fe, while leaving the thermodynamics governing hydrogen mixing in superionic and liquid Fe phases largely unchanged.

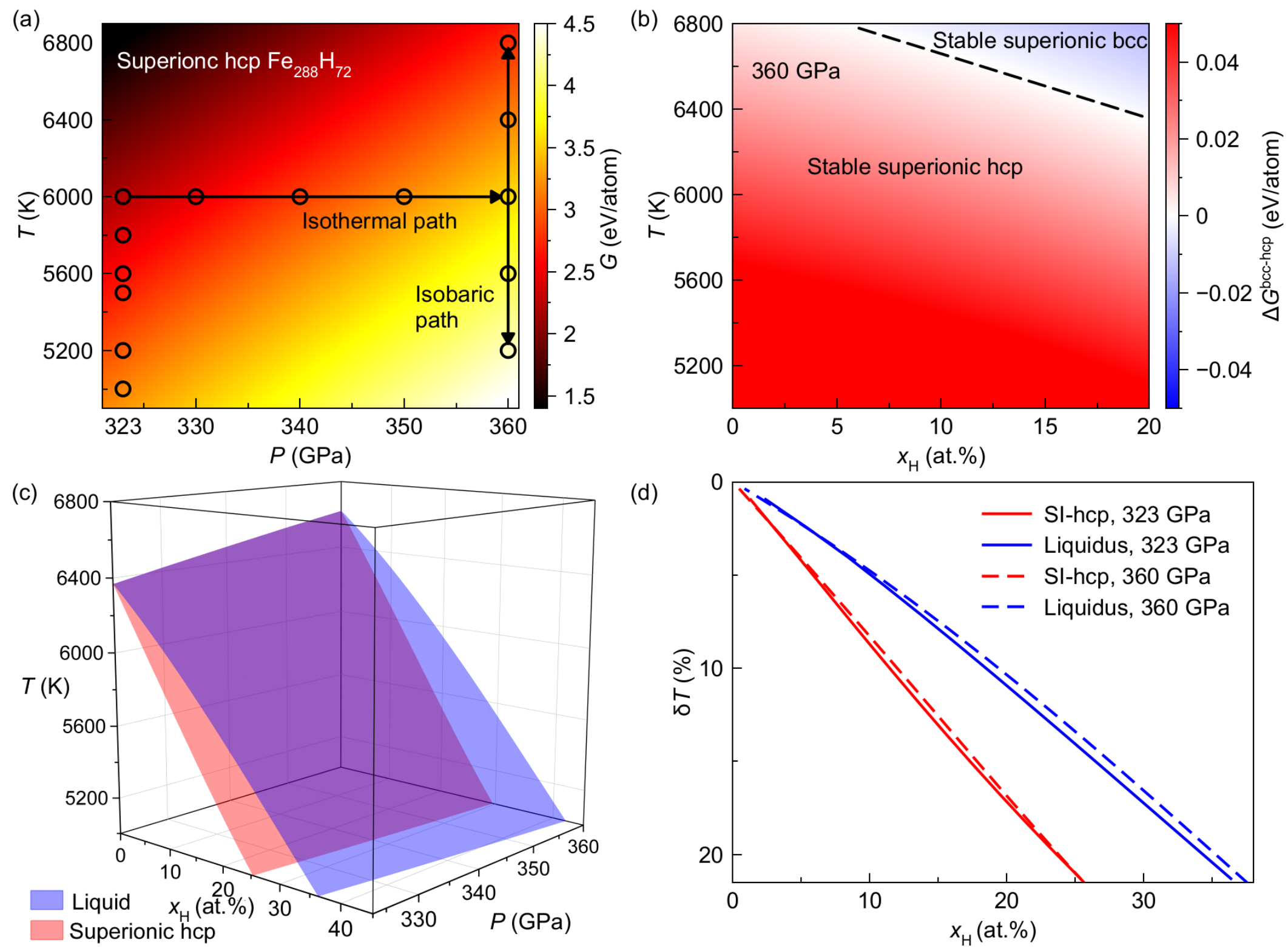

**Figure 2.** ***Ab initio*** **Gibbs free energies and phase diagram of $Fe_{1-x}H_x$ in the pressure range of 323-360 GPa. A.** *Ab initio* Gibbs free energies of superionic hcp $Fe_{288}H_{72}$ phase over 323–360 GPa and 5000–6800 K. Black open symbols represent P-T data used to calculate free energy along isothermal and isobaric paths using Eqs. (5) and (6). The color map shows the free-energy magnitude. **B.** Gibbs free energy differences between superionic bcc and superionic hcp phases, $\Delta G^{bcc-hcp}(x,T)$, as functions of temperature and H content at 360 GPa. The color map represents the free-energy difference, with red and blue regions corresponding to the stability fields of the superionic hcp and bcc phases, respectively. The black dashed line indicates where $\Delta G^{bcc-hcp}$=0. The solid line denotes the liquidus corresponding to coexistence between superionic hcp and liquid phases. **C.** *Ab initio* superionic–liquid phase diagram of the Fe–H system from 323 to 360 GPa. Red and blue surfaces represent the solidus and liquidus, respectively, as functions of pressure. **D.** Solidus and liquidus curves at 323 and 360 GPa plotted using the reduced temperature, $\delta T$, defined as $\delta T = (T_{m,\mathrm{Fe}} - T)/T_{m,\mathrm{Fe}}$, where $T_{m,\mathrm{Fe}}$ is the iron's melting temperature at the corresponding pressure.

### B. Hydrogen distribution in the IC

The superionic–liquid equilibrium condition at the ICB constrains the H composition in both the liquid ($x^{\mathrm{L}}_{\mathrm{ICB}}$) and superionic solid ($x^{\mathrm{SI}}_{\mathrm{ICB}}$), through equality of their chemical potentials, i.e., $\mu^{\mathrm{L}}_{\mathrm{H}}(x^{\mathrm{L}}_{\mathrm{ICB}}, T_{\mathrm{ICB}}, P_{\mathrm{ICB}}) = \mu^{\mathrm{SI}}_{\mathrm{H}}(x^{\mathrm{SI}}_{\mathrm{ICB}}, T_{\mathrm{ICB}}, P_{\mathrm{ICB}})$. More generally, thermodynamic equilibrium also constrains the hydrogen distribution within the IC at different depths. Because hydrogen remains in a liquid-like superionic state throughout the IC, its chemical potential should be spatially uniform, such that

$$\mu^{\mathrm{SI}}_{\mathrm{H}}(x^{\mathrm{SI}}_{\mathrm{H}}(r_{\mathrm{IC}}), T(r_{\mathrm{IC}}), P(r_{\mathrm{IC}})) = \mu^{\mathrm{SI}}_{\mathrm{H}}(x^{\mathrm{SI}}_{\mathrm{ICB}}, T_{\mathrm{ICB}}, P_{\mathrm{ICB}}), \quad (1)$$

where $r_{\mathrm{IC}}$ denotes the radial distance from the center of the IC. This condition provides a strong thermochemical constraint on the hydrogen distribution within the IC. Using the complete free-energy dataset obtained in this work, we explore the hydrogen distribution in the IC under these thermodynamic constraints. Based on the current understanding of the core's thermal structure (33), the IC is nearly isothermal due to its high thermal conductivity (34). We therefore examine the hydrogen chemical potential under isothermal conditions by setting $T(r_{\mathrm{IC}}) = T_{\mathrm{ICB}}$ and adopting $T_{\mathrm{ICB}}$=5500 K (35). Figure 3*A* shows the Gibbs free energy of superionic hcp Fe–H at 5500 K over the pressure range from 330 to 365 GPa, spanning conditions relevant to the IC. We find these free-energy data are well fitted using the Redlich-Kister (RK) model with a fitting error of ~1 meV/atom. This model enables calculation of the hydrogen chemical potential (36) as

$$\mu_{\mathrm{H}}^{\mathrm{SI}}\left(x_{\mathrm{H}}^{\mathrm{SI}}\right) = G^{\mathrm{SI}}\left(x_{\mathrm{H}}^{\mathrm{SI}}\right) + \left(1 - x_{\mathrm{H}}^{\mathrm{SI}}\right)\frac{\partial G^{\mathrm{SI}}(x)}{\partial x}. \quad (2)$$

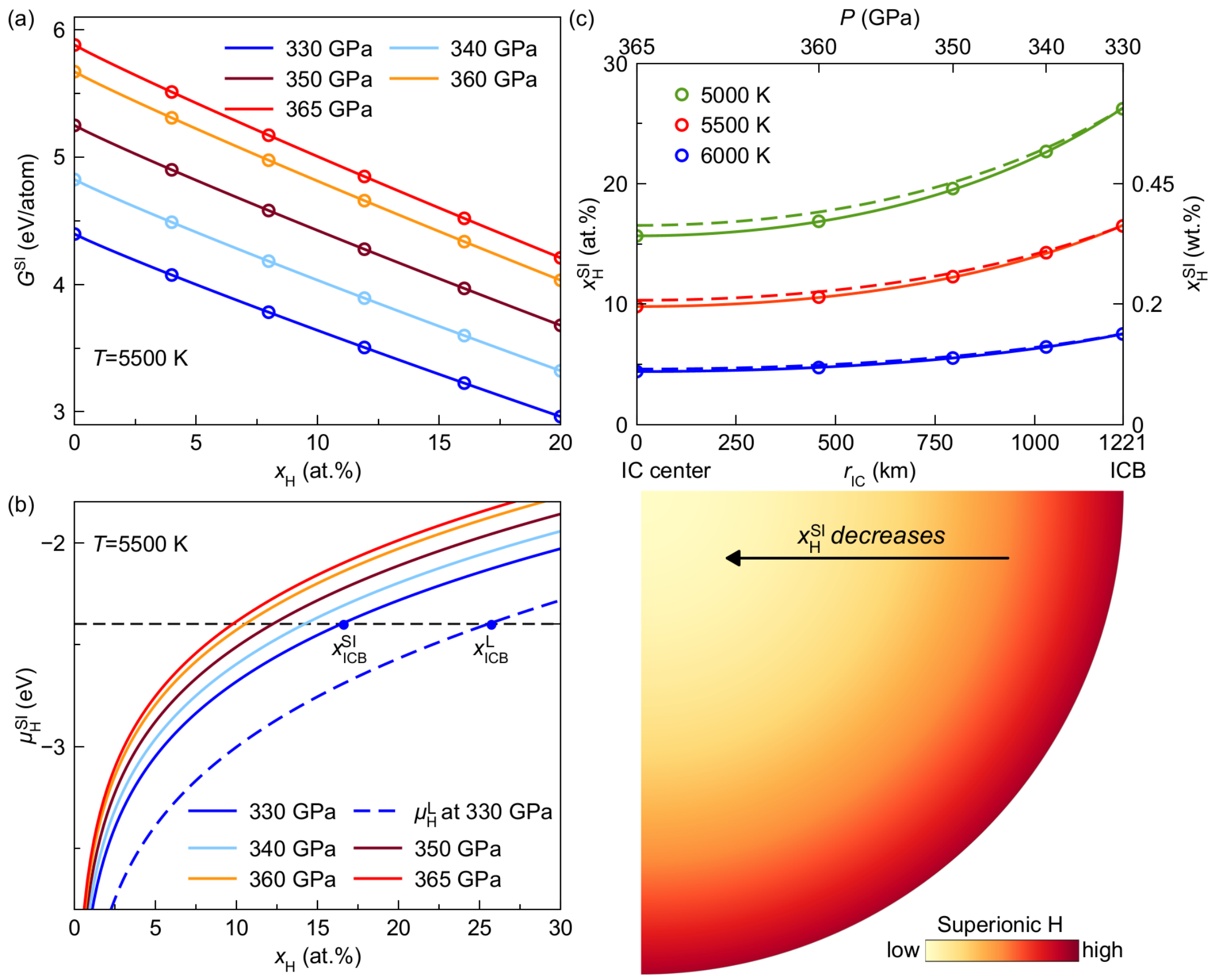

**Figure 3. Radial gradient of superionic hydrogen in Earth's inner core. A.** Gibbs free energy of superionic hcp Fe–H as a function of hydrogen content at 5500 K for different pressures. The solid lines are fits to the data using Eq. (7). **B.** Hydrogen chemical potential in the superionic hcp phase as a function of hydrogen content (solid curves). The dashed blue curve represents the hydrogen chemical potential in the liquid phase at 330 GPa and 5500 K. The dots represent the hydrogen content constrained by the superionic-liquid equilibrium at the ICB under the conditions of 330 GPa and 5500 K. The black dashed line represents a constant hydrogen chemical potential throughout the IC. **C.** Top: Hydrogen concentration as a function of radial distance in the IC. The dashed lines represent the results including the effect of gravity. Bottom: Schematic illustration of the hydrogen concentration gradient within the IC. The color map shows the concentration of superionic hydrogen.

Figure 3*B* shows $\mu_{\mathrm{H}}^{\mathrm{SI}}$ in the superionic hcp phase at 5500 K across a range of pressures. For $T_{\mathrm{ICB}}$=5500 K, the phase diagram at 330 GPa in Fig. 2*C* constrains the solidus $x_{\mathrm{ICB}}^{\mathrm{SI}}$ and liquidus $x_{\mathrm{ICB}}^{\mathrm{L}}$. As a result, the hydrogen chemical potential in the IC is constrained, as indicated by the dashed line in Fig. 3*B*. This constraint leads to varying hydrogen compositions at different pressures, corresponding to different depths within the IC. The resulting depth-dependent hydrogen distribution is presented in Fig. 3*C*. At 5500 K, the H composition is ~16 at.% at ICB ($r_{\mathrm{IC}}$=1221 km). As the radius decreases, the H composition drops sharply, reaching about 12 at.% at $r_{\mathrm{IC}}$=750 km. Toward the IC center, the variation of H composition becomes more gradual. From $r_{\mathrm{IC}}$=250 km to the center, the H composition remains nearly constant. Overall, at 5500 K, the hydrogen concentration decreases by ~7 at.% from the ICB to the IC center. We also perform similar analyses using $T_{\mathrm{ICB}}$=5000 K and 6000 K, respectively (*SI Appendix,* Fig. S16). In all cases, the IC exhibits compositional gradients, although the absolute gradients differ with $T_{\mathrm{ICB}}$. At 5000 K, the difference in hydrogen concentration between the ICB and the center of the IC is ~10 at.%, whereas at 6000 K this difference decreases to ~3 at.%. We also assess the effect of the temperature profile on the chemical gradient by considering an adiabatic limit (*SI Appendix*, Fig. S17). Even under this upper-bound temperature gradient, a hydrogen concentration gradient persists throughout the IC.

Recently, Ganguly discussed the equilibrium radial distribution of light elements in the liquid OC, where the effect of gravity was also considered(37). Here, we evaluate the hydrogen distribution by explicitly accounting for the gravitational field at different depths, under the equilibrium condition:

$$\mu_{\mathrm{H}}^{\mathrm{SI}}\left(x_{\mathrm{H}}^{\mathrm{SI}}(r_{\mathrm{IC}}), T(r_{\mathrm{IC}}), P(r_{\mathrm{IC}})\right) + \Phi(r_{\mathrm{IC}}) = \mu_{\mathrm{H}}^{\mathrm{SI}}\left(x_{\mathrm{ICB}}^{\mathrm{SI}}, T_{\mathrm{ICB}}, P_{\mathrm{ICB}}\right) + \Phi(\mathrm{r}_{\mathrm{ICB}}), \qquad (3)$$

where $\Phi(r_{\mathrm{IC}})$ is the gravitational potential energy, given by $\Phi(r_{\mathrm{IC}}) = \int_0^{r_{IC}} m_{\mathrm{H}} g dr$. As shown in Fig. 3*C*, including gravitational potential energy slightly modifies the hydrogen composition, but the gradient in the IC remains unchanged. This is because the gravitational potential energy difference between the IC center and the ICB is significantly smaller than the corresponding variation in chemical potential (*SI Appendix,* Fig. S18).

## DISCUSSIONS

We have computed the free energies and phase diagram of superionic and liquid Fe–H phases over a broad range of pressure and temperature conditions relevant to the Earth's IC. To compare with previous work, we calculate the molar partition coefficient between the superionic and liquid phases, $D_{\mathrm{mole}}^{\mathrm{S/L}}$, by taking the ratio of the solidus and liquidus compositions directly from the phase diagram in Fig. 2*C*. As shown in Fig. 4*A*, $D_{\mathrm{mole}}^{\mathrm{S/L}}$ ranges from 0.70 at 5000 K to 0.59 at 6000 K at 323 GPa, and from 0.68 at 5200 K to 0.57 at 6400 K at 360 GPa, exhibiting nearly linear dependence on temperature under both pressures. Previous simulations reported different values of $D_{\mathrm{mole}}^{\mathrm{S/L}}$ at various pressures and temperatures. Zhang et al. (30) reported $D_{\mathrm{mole}}^{\mathrm{S/L}}$ values of 0.54 at (5000 K, 250 GPa), 0.56 at (6000 K, 330 GPa), and 0.55 at (6500 K, 360 GPa), in good agreement with our results at (6000 K, 323 GPa). Yuan and Steinle-Neumann (28) reported a value of 0.62 at (6188 K, 330 GPa). In contrast, Liu and Jing (29) reported smaller values, ranging from 0.29 to 0.46 at 6200 K and 330 GPa. We find significant differences in the underlying *ab initio* calculations. Both our work and Zhang et al. (30) employed the PAW16 potential for Fe in DFT calculations, which explicitly includes the contributions from inner-shell *3s* and *3p* electrons. In contrast, the MLPs used in Refs. (28, 29) were trained on DFT data that excluded the *3s* electrons. Moreover, Ref. (28) employed a fixed electronic temperature of 6000 K in the DFT calculations, which may affect the temperature dependence of $D_{\mathrm{mole}}^{\mathrm{S/L}}$. It has been shown that inner-shell electrons and the treatment of electronic temperature significantly impact the thermodynamics of Fe (23, 24, 38). Therefore, regardless of the specific thermodynamic methods used to compute $D_{mole}^{S/L}$, differences in DFT settings that affect Fe thermodynamics may lead to the variations in the resulting hydrogen partition coefficients.

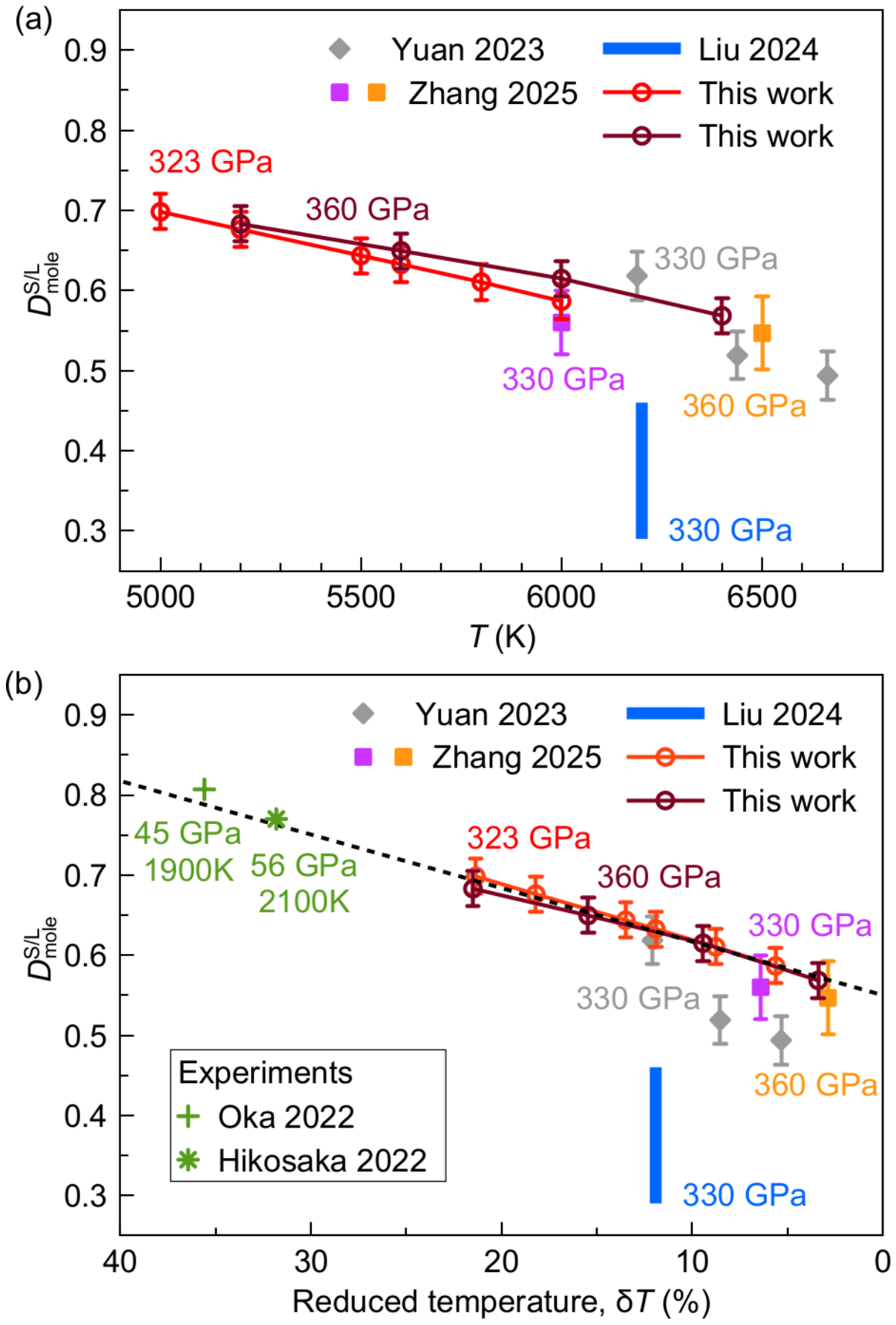

**Figure 4. Hydrogen partition coefficient. A.** Mole partition coefficient between the superionic and liquid phases as a function of temperature. "Yuan 2023", "Liu 2024" and "Zhang 2025" refer to Ref. (28), Ref. (29) and Ref. (30), respectively. **B.** Mole partition coefficient as a function of reduced temperature, $\delta T$. The dashed line represents a linear extrapolation of data from this work at 323 GPa and 360 GPa. "Oka 2022" and "Hikosaka 2022" refer to Ref. (26) and Ref. (27), respectively. For $T_{m,\mathrm{Fe}}$, we adopt values of 6357 K at 323 GPa, 6410 K at 330 GPa, and 6692 K at 360 GPa for DFT calculations including the inner-shell 3*s* and 3*p* electron contribution. For DFT calculations without 3s-electron contribution, a $T_{m,\mathrm{Fe}}$ of 7038 K at 330 GPa is used as the reference (28).

Given that the phase diagrams at 323 GPa and 360 GPa can be scaled by the melting temperature of Fe, as shown in Fig. 2*D*, we replot all the partition coefficients as a function of reduced temperature, $\delta T$, referenced to the corresponding Fe's melting temperature computed using consistent DFT settings, in Fig. 4*B*. We find that the simulation data from our work and the ones from Refs. (28, 30) collapse onto a common trend and exhibit good agreement. The results from Ref. (29) deviate from other data, likely due to the free-energy calculation methodology used in that study. In experiments, Oka et al. (26) reported a partition coefficient of 0.77 by weight fraction, corresponding to 0.81 by molar ratio, at (1900 K, 45 GPa). Hikosaka et al. (27) reported a partition coefficient of 0.7 by weight fraction, or 0.77 by mole ratio, at (2100 K, 56 GPa). Using experimental melting temperatures for Fe ($T_{m,\mathrm{Fe}}$ =2950 K at 45 GPa and 3080 K at 56 GPa) (39), we find that the experimental data, when plotted as a function of $\delta T$, agree remarkably well with the curve extrapolated from our data in Fig. 4*B*, despite the large differences in pressure and absolute temperature.

The consistency between independent DFT calculations and experimental measurements strongly suggests that reduced temperature, $\delta T$ , is the key thermodynamic parameter controlling the hydrogen partition coefficient in the Fe–H system. This behavior can be understood by noting that hydrogen is superionic in the solid phase and fully liquid-like in the melt, leading to similar contributions of the hydrogen subsystem to both phases, as manifested in the H–H PPCF shown in Fig. 1*A*. Thus, the partitioning behavior is primarily governed by the structural and entropic contrast between the solid and liquid Fe sublattices, which is naturally captured by the reduced temperature relative to the melting temperature of Fe.

Because the H partition coefficient depends mainly on reduced temperature and only weakly on pressure (Fig. 4), newly crystallized phase at the ICB is expected to incorporate similar H contents at different stages of IC growth. With continued IC growth, earlier-formed layers become progressively located deeper within the IC. The H contents in these layers should evolve toward the equilibrium radial H gradient (Fig. 3C), driving superionic H outward toward the ICB. Once H reaches the ICB, it preferentially re-partitions into the liquid because $D^{S/L}$<1, further enriching the OC in H. Thus, the present results imply that H exchange between the IC and OC may arise not only from direct partitioning during IC crystallization, but also from the continuous redistribution of superionic H within the growing IC. Both processes may contribute to chemical buoyancy, which is believed to be an important energy source driving the geodynamo (40, 41).

Different H contents between the IC and OC can also contribute significantly to the density contrast at the ICB. With increasing H content in the IC, the density contrast becomes larger (*SI Appendix*, Fig. S19). An H content of 23 atomic percent (at.%), i.e., 0.54 weight percent (wt.%), in the IC is sufficient to reproduce the PREM density contrast by H alone. However, H enrichment also strongly depresses the Fe's melting point. The 23 at.% H content in the IC leads to an ICB temperature of 5150 K and an outer-core H content of 36 at.% (~1 wt.%). The former is too low compared with the lower bound estimated from CMB temperature constraints (1), while the latter exceeds the upper limit of the H content in the OC based on estimates of H partitioning during core-mantle differentiation (14, 42). Therefore, a more plausible scenario likely requires considering the H phase diagram together with other light elements, especially strongly liquid-preferring elements such as O (25, 43) or C (44–46), to jointly account for the density contrast at ICB.

Seismic observations reveal complex anisotropy in the IC, which is commonly attributed to the lattice-preferred orientation of crystalline phases (8, 47). However, the mechanism responsible for the depth-dependent change in anisotropy, which is weak near the ICB but stronger toward the IC center, remains an open question (48, 49). Recent experiments suggest that the anisotropy of Fe-Si-C alloys can be strongly affected by light-element composition (11). Superionic H can also have a complex effect on the anisotropy of Fe-H alloys, with sensitivity to the chemical composition and the geomagnetic field (50). Our predicted radial gradient of H could naturally contribute to the depth-dependent anisotropy of the IC if lower H concentrations, together with increased pressure, enhance the alloy's elastic anisotropy, an important topic for future work. In addition, the bcc phase can exhibit elastic anisotropy distinct from that of hcp Fe under IC conditions (51, 52). Although the superionic bcc phase remains metastable in the Fe-H alloy considered here (Fig. 2B), it may be stabilized by additional alloying elements such as Ni (24) and Si (22). Thus, strong hcp-bcc phase competition in multicomponent Fe alloys may lead to coexistence of these phases in the IC, together with a radial gradient of light elements. Such coupled effects can provide a possible mechanism for depth-dependent IC anisotropy, although a quantitative assessment requires further calculations across the relevant pressure, temperature, and composition ranges.

In summary, we computed comprehensive *ab initio* Gibbs free energies for liquid, superionic hcp, and bcc Fe–H phases and constructed the superionic–liquid phase diagram for pressure–temperature conditions relevant to the IC. Our calculations show that hydrogen can stabilize a superionic bcc phase at sufficiently high temperature and hydrogen content. However, this stability field is superseded by melting, so only the superionic hcp phase coexists with the liquid in the Fe–H system. The superionic–liquid phase diagrams at different inner-core pressures collapse when scaled by the Fe's melting temperature. As a result, the partition coefficient is primarily controlled by the reduced temperature $\delta T$ and shows minimal sensitivity to pressure under this scaling. Moreover, most previous calculations become mutually consistent, and the extrapolated results agree well with low-pressure experimental data when expressed in terms of this scaling relation.

The phase diagram imposes direct thermodynamic constraints on the equilibrium hydrogen contents of coexisting solid and liquid at the ICB. Through thermochemical equilibrium, it further constrains the hydrogen distribution throughout the IC. These constraints predict a radial hydrogen gradient, with hydrogen enriched toward the ICB relative to the center. Our results, therefore, demonstrate that the non-uniform distribution of superionic hydrogen in the IC is a direct consequence of equilibrium thermodynamics. This mechanism may also apply to the distribution of other light elements, influencing core's composition, dynamics and evolution.

## METHODS

### A. *Ab initio* Gibbs free energy calculation

The *ab initio* Gibbs free energies of Fe-H liquid, $G^{\mathrm{L}}$, at 323 GPa were computed using classical-to-*ab initio*

thermodynamic integration (CATI) calculations (20). The free energies of superionic hcp and bcc Fe-H phases were obtained using a workflow (25) that combines solid-liquid coexistence simulations (53) and CATI, utilizing the thermodynamic relation for the superionic-liquid equilibrium as (25)

$$G^{\mathrm{SI}}(x^{\mathrm{SI}}) = G^{\mathrm{L}}(x^{\mathrm{L}}) - (x^{\mathrm{L}} - x^{\mathrm{SI}})\frac{\partial G^{\mathrm{L}}(x^{\mathrm{L}})}{\partial x}, \quad (4)$$

where $x^{\mathrm{L}}$ and $x^{\mathrm{SI}}$ are the equilibrium hydrogen atomic fractions in the liquid and superionic solutions $\mathrm{Fe}_{1-x}\mathrm{H}_x$, respectively, as determined from superionic–liquid coexistence simulations.

The superionic–liquid coexistence simulations were performed using the LAMMPS software package (54). An embedded-atom method (EAM) potential for the Fe–H system was developed to simulate the superionic and liquid Fe-H phases using an iterative algorithm (55) (see *SI Appendix*, Note S1 and Fig. S1 for details). The superionic–liquid coexistence model contained 36,960 Fe atoms and up to 11,208 H atoms. Temperature and pressure were controlled using the Nosé–Hoover thermostat and barostat with a damping time of τ=0.1 ps. The simulations were carried out with a time step of 0.25 fs for 2 ns, sufficient to equilibrate the system and ensure convergence of the superionic–liquid equilibrium.

*Ab initio* molecular dynamics (AIMD) simulations were performed using the Vienna ab initio simulation package (VASP) (56). The projected augmented wave (PAW) method (57) was used to describe electron-ion interactions, and the generalized gradient approximation (GGA) in the Perdew-Burke-Ernzerhof (PBE) (58) form was used for the exchange-correlation functional. We employed the Mermin functional (59, 60) to include the electronic entropy, with the electronic temperature set equal to the ionic temperature in AIMD simulations. Supercells containing 250–360 atoms were used to avoid the finite-size effects in the CATI calculation. The time step was set to 0.5 fs for AIMD and 1 fs for CATI simulations. For Fe, the PAW potential with $3d^7 4s^1$ valence electrons was used in AIMD and CATI simulations, while the PAW potential with $3s^2 3p^6 3d^7 4s^1$ valence electrons was used in free energy perturbation (61) (FEP) calculations which is necessary to incorporate the inner-shell electron contributions to the free energy (20, 24). PAW potential with $1s^1$ valence electrons was employed for H. A plane-wave cutoff energy of 400 eV and $\Gamma$-point sampling were applied in both AIMD and CATI simulations. A higher ENCUT of 750 eV and a denser k-point mesh of $2 \times 2 \times 2$ were used in FEP calculations. The lattice parameters of liquid, superionic hcp and superionic bcc phases were adjusted at each temperature and composition to maintain the pressure fluctuations within ±0.5 GPa over 5 ps of NVT-AIMD simulations. Enthalpy data were collected from AIMD runs lasting more than 8 ps.

### B. Thermodynamical models

The free energy computed at ($P_0$ =323 GPa, $T_0$ =6000 K) were extended to other ( $P, T$ ) conditions via a few thermodynamic relations. The free energy was computed along the isothermal line for different pressures via direct integration of equation of states $V(T, P)$ under constant $T_0$ as

$$G(T_0, P) = G(T_0, P_0) + \int_{P_0}^{P} V(T_0, P)dP. \quad (5)$$

Here, $V(T, P)$ was computed by AIMD and fitted via the third-order Birch–Murnaghan equation of state (62, 63). More details are in *SI Appendix*, Note S3. The Gibbs–Helmholtz equation was employed to compute the free energy at different $T$ along the isobaric line under constant $P_0$ as

$$G(T, P_0) = \frac{G(T_0, P_0)}{T_0}T - T\int_{T_0}^{T}\frac{H(T, P_0)}{T^2}dT, \quad (6)$$

where $H(T, P_0)$ is the enthalpy. The composition-dependent Gibbs free energy for $\mathrm{Fe}_{1-\mathrm{x}}\mathrm{H}_{\mathrm{x}}$ is fitted by Redlich-Kister (RK) expansion [32] as

$$G(x) = G_{Fe} + ax + k_B T[(1-x)\ln(1-x) + x\ln x] + x(1-x)\sum_{n=0}(1-2x)^n L_n, \quad (7)$$

where $G_{Fe}$ is the Gibbs free energy of pure Fe. a and $L_n$ are the fitting parameters.

## ACKNOWLEDGMENTS

Work at Xiamen University was supported by the National Natural Science Foundation of China (Grants T2422016, 42374108 and 12374015). The work at Columbia University was supported by the Gordon and Betty Moore Foundation Award GBMF12801 (doi.org/10.37807/GBMF12801). Shaorong Fang and Tianfu Wu from the Information and Network Center of Xiamen University are acknowledged for their help with Graphics Processing Unit computing. Some ab initio simulations were performed on Bridges-2 system at PSC, the Anvil system at Purdue University, the Expanse system at SDSC, and the Delta system at NCSA through allocation DMR180081 from the Advanced Cyberinfrastructure Coordination Ecosystem: Services & Support (ACCESS) program, which is supported by NSF Grants No. 2138259, No. 2138286, No. 2138307, No. 2137603, and No. 2138296. The supercomputing time were also supported by the Opening Project of the Joint Laboratory for Planetary Science and Supercomputing, Research Center for Planetary Science, and the National Supercomputing Center in Chengdu (Grants No. CSYYGS-QT-2024-15).

---